\documentclass[%
twocolumn,
prl,
showpacs,preprintnumbers,
nofootinbib,
amsmath,amsthm,amssymb,
]{revtex4}

\usepackage{bbm}
\usepackage{slashed}
\usepackage{color}

\usepackage{epsfig,graphicx,bm,color} 
\usepackage{dcolumn}
\usepackage{bm}
\usepackage{hyperref}
\usepackage{color}
\usepackage{breakurl}
\usepackage{soul} 

\newcommand{\ds}{{\sf DarkSUSY}}
\newcommand{\be}{\begin{equation}}
\newcommand{\ee}{\end{equation}}
\newcommand{\bea}{\begin{eqnarray}}
\newcommand{\eea}{\end{eqnarray}}

\newcommand{\lsim}{\mathrel{\mathop{\kern 0pt \rlap
  {\raise.2ex\hbox{$<$}}}
  \lower.9ex\hbox{\kern-.190em $\sim$}}}
\newcommand{\gsim}{\mathrel{\mathop{\kern 0pt \rlap
  {\raise.2ex\hbox{$>$}}}
  \lower.9ex\hbox{\kern-.190em $\sim$}}}

\begin{document}

\title{Significant Enhancement of Neutralino Dark Matter Annihilation\\ from Electroweak Bremsstrahlung}

\author{Torsten Bringmann}
\email{Torsten.Bringmann@fys.uio.no}
\affiliation{Department of Physics, University of Oslo, Box 1048 NO-0316 Oslo, Norway}
\affiliation{\mbox{{II.} Institute for Theoretical Physics, University of Hamburg,
Luruper Chaussee 149, DE-22761 Hamburg, Germany}}

\author{Francesca Calore}
\email{f.calore@uva.nl}
\affiliation{GRAPPA Institute, University of Amsterdam, Science Park 904, 1090 GL Amsterdam, The Netherlands}
\affiliation{\mbox{{II.} Institute for Theoretical Physics, University of Hamburg,
Luruper Chaussee 149, DE-22761 Hamburg, Germany}}

\date{19 February 2014}

\begin{abstract}
Indirect searches for the cosmological dark matter have become ever more competitive during the past years. Here, we report the first full calculation of leading electroweak corrections to the annihilation rate of supersymmetric neutralino dark matter. We find that these corrections can be huge, partially due to contributions that have been overlooked so far.
Our results imply a significantly enhanced discovery potential of this well motivated dark matter candidate with current and upcoming cosmic ray experiments, in particular for gamma rays and models with somewhat small annihilation rates at tree level. 
\end{abstract}


\maketitle


\paragraph{Introduction.}Thermally produced weakly interacting massive particles (WIMPs) constitute a prime candidate for the so far unexplained  cosmological dark matter (DM), with the lightest supersymmetric neutralino, henceforth denoted by $\chi$, being one of the best motivated and most often studied examples \cite{Jungman:1995df}. Indirect searches for WIMP DM, looking for DM annihilation products in the galactic halo or at cosmological distances, are putting ever more stringent constraints on the annihilation rate, and operating or upcoming experiments like Fermi \cite{Atwood:2009ez}, AMS \cite{Battiston:1999yb} or CTA \cite{Doro:2012xx} are expected to explore a large part of the remaining parameter space of viable models. Because of their outstanding  constraining and signal identification power, gamma rays have been argued to be the 'golden channel' of indirect searches (see Ref.~\cite{Bringmann:2012ez} for a recent review) and this is what we will mostly focus on here.

In view of  the small galactic velocities $v\!\sim\!10^{-3}$ of DM particles, there is little hope for such experiments to be sensitive to anything but  $s$-wave annihilation. For Majorana particles like the neutralino, however, the annihilation into fermionic two-body states $\bar f f$ is  helicity suppressed in the $v\rightarrow0$ limit, by a factor of $m_f^2/m_\chi^2$, because the incoming DM particle pair forms a $J=0$ state \cite{Goldberg:1983nd}.  This suppression can be avoided in the presence of an additional photon in the final state, in particular if emitted from a virtual sfermion with a mass close to $m_\chi$ \cite{Bergstrom:1989jr,Flores:1989ru}.
The radiative 'correction'  thus becomes parametrically as large as $\alpha_{\rm em}/\pi\, (m_\chi^2/m_f^2)$, implying an annihilation into $\bar f f\gamma$ final states at a rate up to several orders of magnitude above the tree-level result. The same mechanism works of course also for the emission of other vector bosons and has thus been intensely studied in the context of indirect DM searches not only for photons \cite{Baltz:2002we,Bringmann:2007nk,Bergstrom:2008gr,Bringmann:2012vr,Bergstrom:2012bd,Shakya:2012fj,Garny:2013ama,Tomar:2013zoa,Toma:2013bka,Giacchino:2013bta}, but also for electroweak gauge bosons \cite{Bell:2010ei, Bell:2011eu,Garny:2011cj,Ciafaloni:2011sa,Bell:2011if,Ciafaloni:2011gv,Ciafaloni:2012gs,Fukushima:2012sp,Shudo:2013lca} and gluons \cite{Asano:2011ik,Garny:2012eb} (see Ref.~\cite{Garny:2011ii} for a recent general analysis).

Phenomenologically, the most important characteristics of photon 'internal bremsstrahlung' (IB)  \cite{Bringmann:2007nk} are the associated pronounced spectral features in gamma rays that can appear near the highest kinematically accessible energies (notably not only for fermion, but also for $W^+W^-$ final states \cite{Bergstrom:2005ss,Garcia-Cely:2013zga}). 
Such gamma-ray features can help tremendously to both detect a DM signal and to distinguish it from astrophysical backgrounds \cite{Bringmann:2011ye,Bringmann:2012vr}. 

Electroweak or strong gauge boson IB, on the other hand,  generically proceeds at considerably higher rates due to the larger coupling strength involved. As a result of the decay and fragmentation of the additional gauge boson, however, it affects the gamma-ray spectrum only at energies much below the DM mass and does not introduce any pronounced spectral features. 
The potentially large enhancement can still be very important for photon (or other cosmic ray) counting experiments with low energy thresholds. It was also pointed out that electroweak IB can be phenomenologically important even when not lifting the helicity suppression, like for final state radiation, because it may significantly alter the composition of the DM-induced cosmic-ray spectrum  \cite{Kachelriess:2007aj,Kachelriess:2009zy,2010PhRvD..82d3512C,Ciafaloni:2010ti}.

Encouraged by the importance of an additional $W/Z$ boson in fermion final states as found in previous studies, we present here the first \emph{fully general} calculation for neutralino DM, keeping not only all relevant diagrams but also the full mass dependence of both fermions and gauge bosons. We identify  new enhancement mechanisms and show that model-independent results \cite{Ciafaloni:2010ti,Cirelli:2010xx}, though being quite popular, are in general not sufficient to produce realistic estimates of DM indirect detection prospects.

\paragraph{Neutralino annihilation into fermions and $Z/W^\pm$.}

\begin{figure*}[!t]
\begin{center}
\includegraphics[width=1.9\columnwidth,clip]{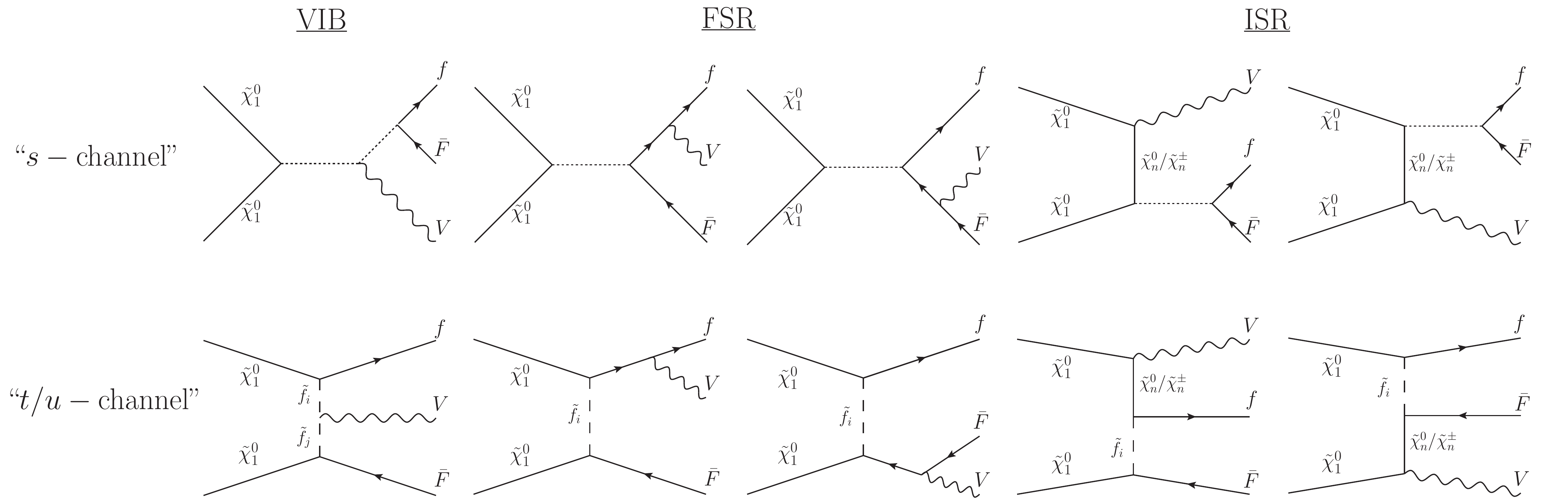}  
\caption{Condensed representation of all Feynman diagrams for  $\chi\chi\to\bar{F} fV$, where $F=f$ for $Z$-boson emission ($V\!=\!Z$) and $(F,f)$ are the two components of an $SU(2)_{L}$ doublet for $W$-boson emission ($V\!=\!W^\pm$). Dotted lines indicate scalar ($A$, $h$, $H$, $H^{\pm}$) or vector ($Z,W^{\pm}$) mediators.
}
\label{fig:diagrams}  
\end{center}
\end{figure*}

In the minimal supersymmetric standard model (MSSM), the lightest neutralino is a linear combination of the neutral superpartners of the gauge and Higgs fields,
$\chi\equiv\chi^0_{1} = N_{11}\, \tilde{B} + N_{12}\,\tilde{W}^3 + N_{13}\, \tilde{H}^0_1 + N_{14}\,\tilde{H}^0_2 
$,
with $Z_{g} \equiv |N_{11}|^2 + |N_{12}|^2$  indicating its gaugino fraction.
We collect in Fig.~\ref{fig:diagrams}, in a very condensed form, all  Feynman diagrams
 that contribute to neutralino annihilation into two fermions and one $SU(2)$ gauge boson,
   in total 46 for $W^\pm$ and 88 for $Z$ emission, where dotted lines indicate scalar  ($A$, $h$, $H$ and  $H^{\pm}$) or vector ($Z,W^{\pm}$) mediators. Diagrams in the first (second) row derive from tree-level $s$-($t$-)channel diagrams; $u$-channel derivates  are not shown explicitly and for  $v\rightarrow0$ simply result in the \emph{same} amplitudes as the $t$-channel case. For each tree-level topology, the diagrams can further loosely be classified by whether the electroweak gauge boson is radiated from the virtual particle that appears at tree-level (Virtual Internal Bremsstrahlung (VIB), first column),  one of the final fermions (Final State Radiation (FSR),  second and third column), or by one of the initial neutralino legs (Initial State Radiation (ISR), column 4 and 5).

We generate those diagrams with {\sf FeynArts} \cite{FeynArts}, modifying the supplied generic MSSM model file such as to agree with the conventions adopted in \ds\ \cite{DS}, and use {\sf FeynCalc} \cite{Mertig:1990an} to calculate the amplitudes. 
In order to obtain tractable analytic expressions we  restrict ourselves to the \mbox{$v\rightarrow0$} limit, in which only the singlet state ($J^P\!=\!0^-$) of the neutralino pair contributes; in practice, we replace the pair of external Majorana spinors in the amplitude by the Lorentz-invariant projector 
$P_{\,^{1}\!S_{0}} = \frac{\gamma_5}{\sqrt{2}} \left(m_{\chi} - \slashed{p}/2\right)$,
where $p$ is the total momentum of the system. 
In the next step, we adapt the \textsl{helicity amplitude} method which has proven very useful for the numerical analysis of neutralino annihilation into two-body final states \cite{Edsjo:1997bg,Edsjo:2003us} to our case. Concretely, we compute the total amplitude squared, summed over final and averaged over initial spin degrees of freedom, as
\be
\overline{\left|\mathcal{M}\right|}^2
=\frac{1}{4} \sum_{h, \lambda} \,\Big|\!\sum_{\rm diag.}\! \mathcal{M}^{(h,\lambda)}_{\chi \chi \rightarrow \bar{F} f V}\Big|^2\,.
\ee
Here, $h$ refers to the helicity of the fermion-antifermion pair in its rest-frame (where $F$ equals $f$ for $V\!=\!Z$, while for $V\!=\!W^\pm$ it is given by its  $SU(2)_{L}$ partner),  $\lambda$ is the polarization state of $V$ in the same system and the inner sum runs over all diagrams of Fig.~\ref{fig:diagrams} (for more technical details, see Ref.~\cite{fullanalysis}).
Finally, the cross section is obtained by integration over the three-body phase space:
\begin{equation}
\label{sdiff}
\frac{d (\sigma v)}{dE_1 dE_2} = \frac{1}{16 \, m_{\chi}^{2}} \, \frac{1}{(2 \pi)^{3} } \overline{\left|\mathcal{M}\right|}^2\,,
\end{equation}
where $E_1$ and $E_2$ are the center-of-mass system (CMS) energies of any two final state particles.

As a cross-check of our results, we analytically reproduced the differential cross section for $\chi\chi\rightarrow\bar f f\gamma$ \cite{Bringmann:2007nk}, for all diagrams and $m_f\neq0$, by taking the appropriate limit of our expressions for $ \bar f f Z$ final states. We also find exact agreement for $\bar f fZ$ and $\bar F fW$ final states in the limits considered in Ref.~\cite{Garny:2011ii}, which extends earlier partial results for those processes \cite{Bell:2010ei,Bell:2011eu,Garny:2011cj}, i.e.~$m_f\equiv0$ and $\chi$ being a pure Bino or Higgsino.
Compared to those references, however, our expressions are considerably more complex because we keep the full mass-dependence as well as the much larger number of diagrams that appear for mixed neutralinos. All matrix elements and cross sections have been implemented in \ds\ and will be available with the next public release.

\paragraph{Spectra of stable annihilation products.}

Integration of Eq.~(\ref{sdiff}) directly gives the spectrum of any of the final state particles $p$ -- which, however, mostly fragment or decay. The spectrum of a potentially observable stable particle $P$, resulting from a given annihilation channel $\chi\chi\rightarrow\bar F f V$ and normalized to the {\it total} tree-level annihilation rate $\sigma v_{0}^{\rm tot}$, can  be written as 
\begin{equation}
\label{spectrum}
\frac{dN_P^{\bar{F} f V}} {dE_P} =\sum_{p= F, f,V} \int^{E_{p}^{\rm max}}_{E_{p}^{\rm min}}  \frac{1}{2}\frac{dN^{\bar{p} p \rightarrow P+ X}_P} {dE_P}  \frac{dN^{\bar{F} f V}_p} {dE_p} \, dE_p\, ,
\end{equation}
where 
\begin{equation}
\frac{dN^{\bar{F} f V}_p} {dE_{p}} = \frac{1}{{\sigma v}_{0}^{\rm tot}} \int^{E_{p'}^{\rm max}(E_{p})}_{E_{p'}^{\rm min}(E_{p})} \frac{d (\sigma v)}{dE_{p}  dE_{p'}} dE_{p'}
\label{eq:spectrumffV}
\end{equation}
and ${dN^{\bar{p} p \rightarrow P+ X}_P}/{dE_P}$ is the Monte Carlo (MC) simulated number of stable particles $P$ resulting from the inclusive process $\bar{p} p\rightarrow P + X$ (evaluated for a CMS energy of $2 E_p$). Those MC spectra are implemented in  \ds, based on a large number of {\sf PYTHIA} \cite{Pythia64} runs.
They have a strong and characteristic dependence on the fractional CMS energy $x_{P} \equiv E_P / E_p $ carried by the stable particle $P$ (while the  dependence on the energy scale $E_p$ itself is rather weak, at least for $E_p\gg m_p$). 
The total photon yield from the $SU(2)$ corrections reported here, e.g., is thus given  by the sum over all  final states $\bar{F} fV$:
\begin{equation}
\frac{dN_{\gamma}^{SU(2)}} {dE_{\gamma}}(E_{\gamma}) = \sum_{i\in \{\bar{F} fV\}}\frac{dN_{\gamma}^i} {dE_{\gamma}}(E_{\gamma})\,.
\label{eq:finalSpect}
\end{equation}

\paragraph{Parameter scan and cross section results.---}

\begin{figure}[t!]
\includegraphics[width= 0.9 \columnwidth,clip]{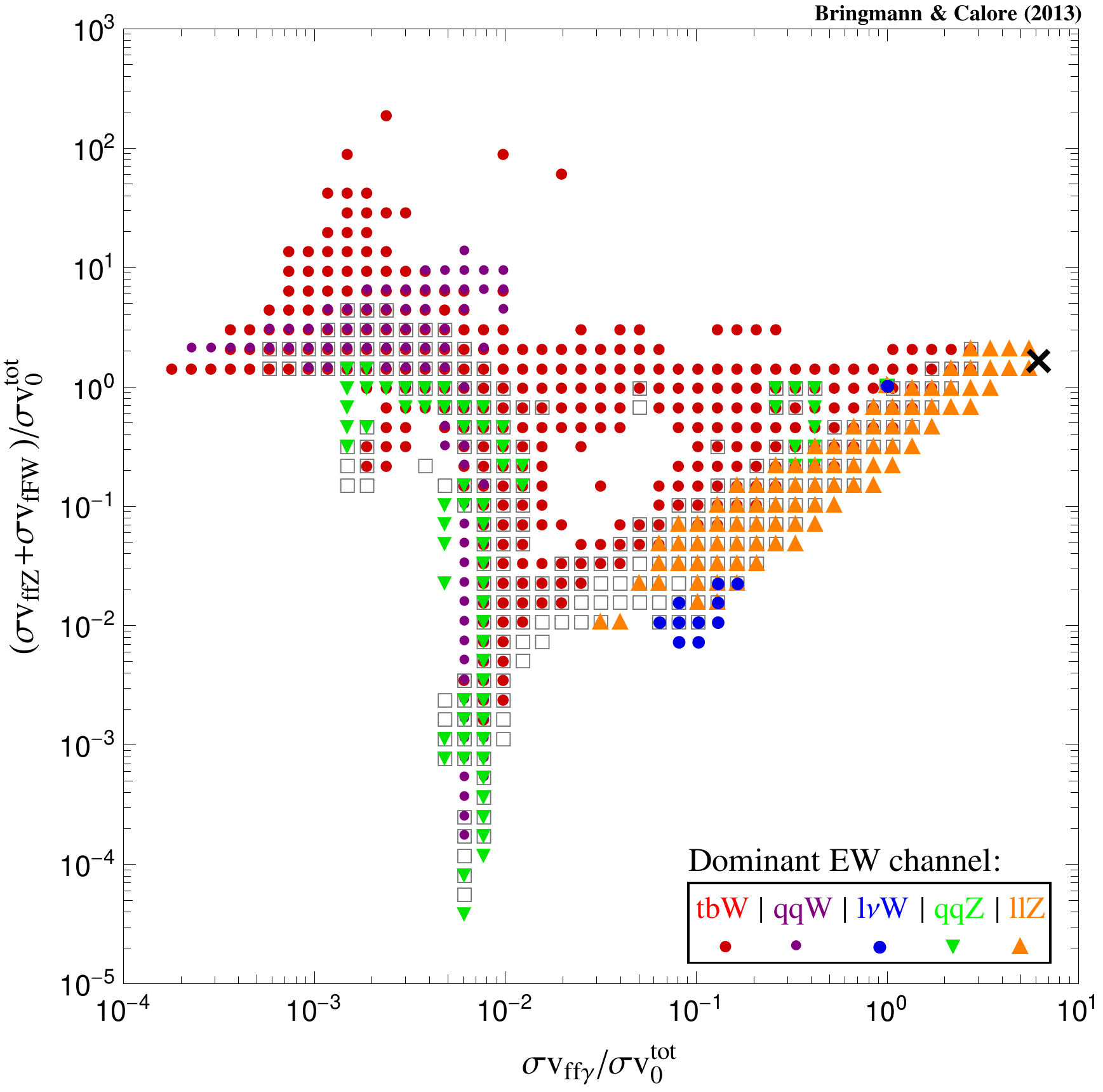}  
  \caption{Total annihilation cross section of thermally produced neutralino DM, in the cMSSM and MSSM-7, into any fermion pair and a
  photon (with $E_\gamma>0.5\,m_\chi$) or an electroweak gauge boson, divided by the total cross section at tree level. The dominant $SU(2)$ annihilation channel is also indicated, with $q$ being any quark but the top (note that $\bar t tZ$ never dominates; empty squares correspond to models where no single channel contributes more than 50\%). The black cross marks the example model given in  Tab.~\ref{tab_susy}.}
  \label{fig:sigmascan} 
\end{figure}

In order to illustrate the potential importance of the processes calculated here, we consider an extensive scan over the parameter space of the constrained MSSM (cMSSM) and a phenomenological MSSM-7 (see Ref.~\cite{Bringmann:2012ez} for details), where we only keep those $\sim4\cdot10^5$ models with a neutralino relic density within the $3\sigma$ estimate of Planck \cite{Ade:2013lta}. In Fig.~\ref{fig:sigmascan}, we compare the total annihilation rate into three-body final states including a fermion pair and an electroweak gauge boson with that for $\bar f f\gamma$ final states,
indicating also the dominant $SU(2)$ annihilation channel.\footnote{
Integrating Eq.~(\ref{sdiff})  leads to a logarithmic divergence for $E_V\!\rightarrow\!0$, which would be   cancelled by the contribution from loop-corrections to the two-body annihilation process. Given the large values of  $m_Z$ and $m_W$, however, we only expect a very small impact of that effect in our case.
Low-energy photons from $\bar f f\gamma$, on the other hand, are phenomenologically not important as they are always vastly dominated by secondary photons at tree-level  \cite{Bringmann:2007nk}. We thus simply refer here to $(\sigma v)_{\bar f f\gamma}$ as the {\it cross section into high-energy photons only}, with  $E_\gamma>0.5\,m_\chi$.
}
 
The first  thing to notice is that electroweak corrections can be quite a bit larger than electromagnetic corrections, and that the correlation between those two is generally not all too pronounced. The reason for this can be found in the much larger number of diagrams that are involved, which results in a rich and complex phenomenology. A full discussion is beyond the scope of this Letter and will be presented elsewhere \cite{fullanalysis}, but we will in the following point out the main enhancement  mechanisms that result in the structure visible in Fig.~\ref{fig:sigmascan}.

The only case where there actually {\it is} a clear correlation between the two IB contributions is the pronounced orange strip with $\ell^+\ell^-Z$ final states dominating the electroweak IB. Those models -- as well as those dominated by $\ell^\pm\nu W^\mp$ -- mostly lie in the cMSSM stau co-annihilation region, where the mass degeneracy between sleptons and the neutralino results in a well-known strong enhancement of $t$-channel VIB diagrams \cite{Bringmann:2007nk,Bergstrom:2008gr,Ciafaloni:2011sa,Bell:2011if,Garny:2011ii} (somewhat suppressed in the electroweak case due to destructive interference with $s$-channel diagrams). The vertical strip with dominantly  $\bar q qZ$ final states mostly corresponds to Binos resonantly annihilating via $s$-channel $A$ exchange. FSR cannot lift the helicity suppression here, which explains the smallness of the $U(1)$ contributions; moving up the strip, the neutralinos feature an increasing Higgsino fraction -- which opens up the ISR channels and thus potentially large  enhancement factors due to a lifting of the helicity suppression \cite{Ciafaloni:2011gv}. 

The largest enhancements appear for $tbW$ final states (note that $qqW$ only dominates for $2m_\chi\lesssim m_t+m_W$), due to various reasons. One of the strongest enhancement mechanisms that we identified is a threshold effect \cite{Chen:1998dp,Yaguna:2010hn,Honorez:2010re} for $m_\chi\lesssim m_t$, where $\bar t t$ final states are kinematically not allowed or strongly suppressed. Resonances  can also play an important role for some models, as well as the fact that the $s$-channel amplitude is directly proportional to the large top mass.

\paragraph{Implications for DM searches.---}

\begin{figure}[t!]
\includegraphics[width= 0.9 \columnwidth,clip]{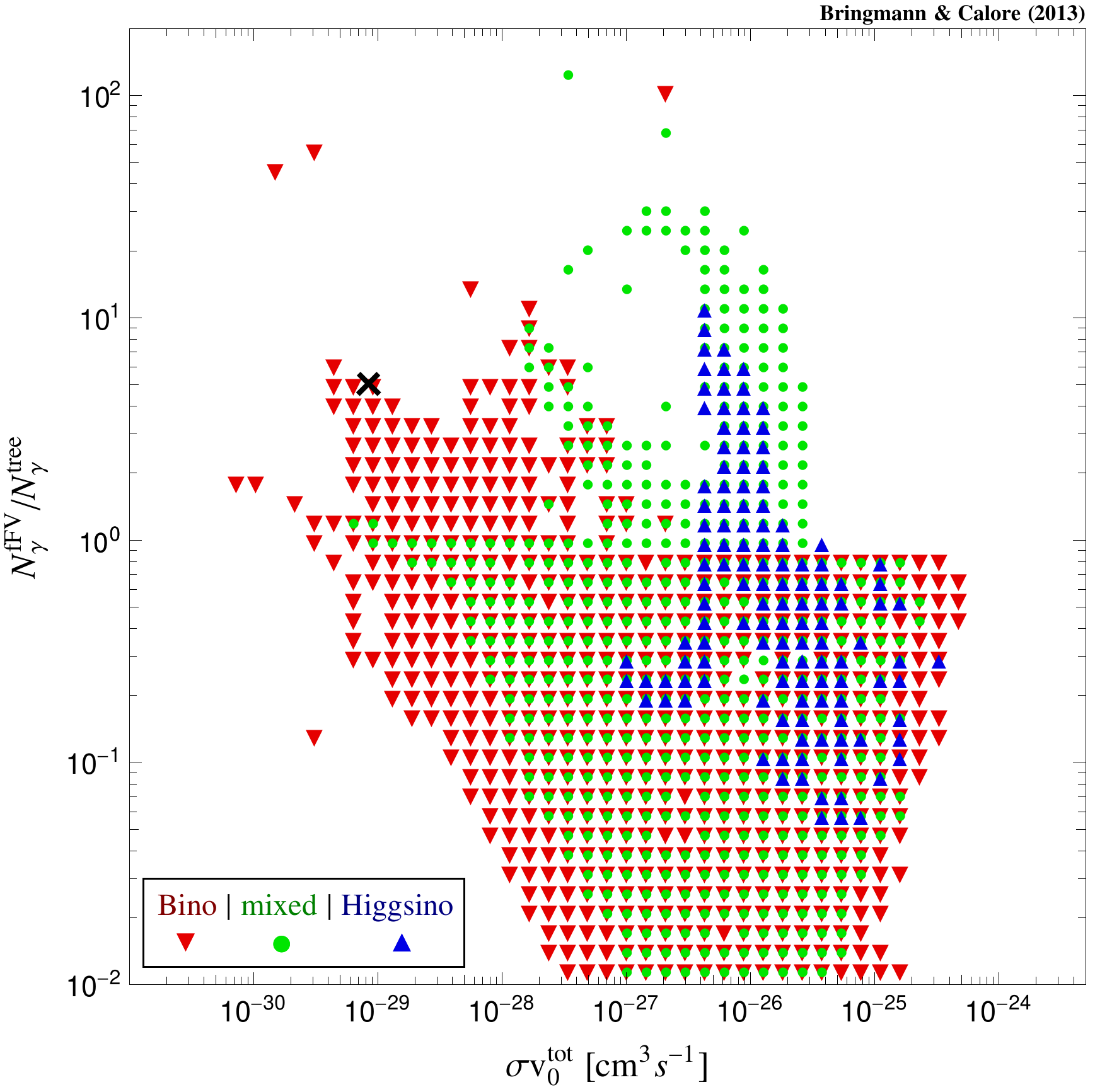}  
  \caption{Photon yield enhancement (for $E_\gamma>100$\,MeV) from neutralino DM annihilation due to electroweak radiative corrections vs.~the full annihilation rate $\sigma v_0^{\rm tot}$ at tree level (in the cMSSM and MSSM-7). Symbols indicate models where the neutralino is mostly Bino ($Z_g\!>0\!.99$), Higgsino ($Z_g\!<\!0.01$) or mixed ($0.01\!\leq\! Z_g\!\leq\!0.99$). The black cross marks the example model given in  Tab.~\ref{tab_susy}.}
  \label{fig:Ngamma} 
\end{figure}

For indirect DM detection, a more suitable measure for the impact of the radiative corrections reported here is the enhancement  of cosmic ray yields at low energies. 
In Fig.~\ref{fig:Ngamma} we  show this quantity  for photons with $E_\gamma>$100\,MeV (roughly  the threshold of Fermi). The photon count can be enhanced by up to two orders of magnitude, in particular for models where the total annihilation rate at tree-level is smaller than the 'canonical' value of  $\langle \sigma v \rangle_{\rm therm} \equiv 3\times 10^{-26}\rm \ cm^3 s^{-1}$ for thermally produced DM. Comparing this to the model-independent limits derived from 3 years of  dwarf galaxy observations  \cite{GeringerSameth:2011iw}, we find that even some thermally produced neutralinos as heavy as several hundred GeV could eventually be probed by Fermi once electroweak corrections are taken into account.\footnote{
For neutralino models that could explain the monochromatic $\sim$130 GeV signal around the galactic center \cite{Bringmann:2012vr,Weniger:2012tx} in terms of photon IB, on the other hand, the low-energy photon yield is never enhanced by more than a factor of $\sim$2 (c.f.~Fig.~4 in Ref.~\cite{Bringmann:2012ez}). 
}

Even when increasing the energy threshold to $E_{\rm thr}\sim100$\,GeV, as is typical for currently operating Air Cherenkov telescopes, the resulting enhancement in the photon flux is very sizable -- not the least because particularly large effects can be found for DM masses in the TeV range (in which case one has $m_\chi\gg E_{\rm thr}$, implying that the photon yield enhancement is very similar to the one  shown in Fig.~\ref{fig:Ngamma} for $E_{\rm thr}=0.1$\,GeV). Given its lower targeted threshold, radiative corrections will  obviously be even more important when interpreting the results of DM searches with CTA. 
While we have limited the above discussion to gamma rays, we note that corresponding considerations apply to other cosmic ray species  \cite{fullanalysis}. In particular, electroweak corrections can completely dominate the neutrino signal \cite{Fukushima:2012sp,Shudo:2013lca} and prospects for indirect DM searches with antiprotons or positrons through the AMS experiment  can be improved significantly (especially, as expected \cite{Evoli:2011id},  for light neutralinos). 

It is worth mentioning that elastic scattering rates are rather uncorrelated with the three-body cross sections calculated here; in particular, we checked that including the most recent direct detection bounds \cite{Akerib:2013tjd} does not qualitatively change Figs.~\ref{fig:sigmascan} or \ref{fig:Ngamma}. While such a correlation is also absent at tree-level  \cite{Bergstrom:2010gh}, it may in contrast be quite strong when considering instead {\it electromagnetic} corrections in models with  very small mass differences between squarks and neutralinos  \cite{Garny:2013ama} (because in this case the zero photon mass and the absence of destructively interfering $s$-channel diagrams lead to an unsuppressed enhancement).

\begin{figure}[t!]
\includegraphics[width= 0.9 \columnwidth,clip]{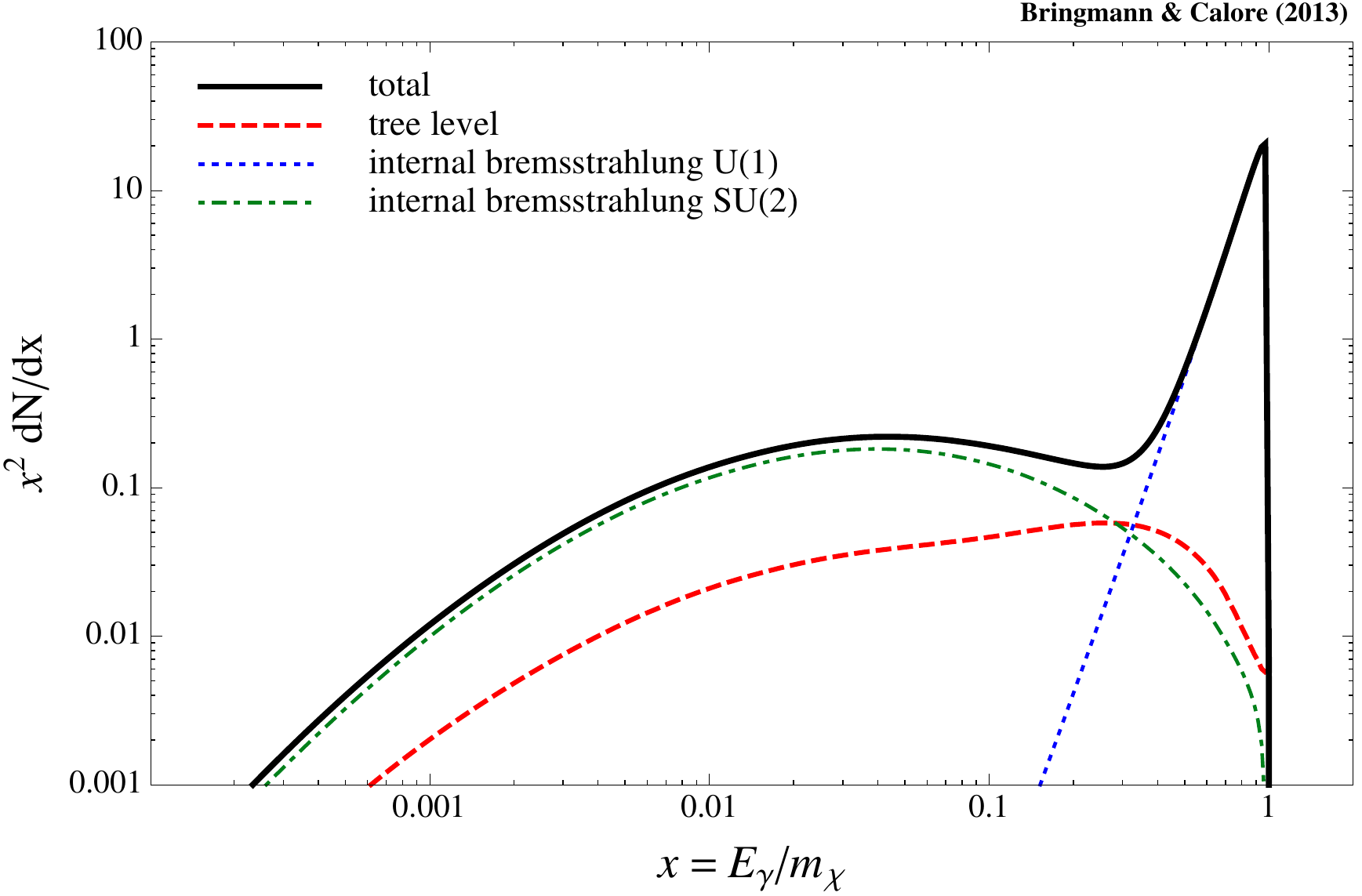}  
  \caption{Gamma-ray spectrum from neutralino DM annihilation for the example model of Table \ref{tab_susy} (solid line). Shown separately is the tree-level annihilation spectrum (dashed line), the spectrum from electroweak corrections only (computed in this Letter, dashed-dotted line) and from  electromagnetic corrections only (following Ref.~\cite{Bringmann:2007nk}, dotted line). }
  \label{fig:spectrum} 
\end{figure}

Last but not least, we illustrate in Fig.~\ref{fig:spectrum} how the {\it spectral shape} of gamma rays from DM annihilation can be distorted when including electroweak corrections.
The example shown here corresponds to a typical Bino-like cMSSM model in the $\tilde\tau$-coannihilation region, with model parameters as specified in Tab.~\ref{tab_susy} and satisfying current LHC limits on the squark and gluino masses \cite{ATLAS_cMSSM,CMS_cMSSM}. 
As expected, the effect of electroweak corrections results in a rather feature-less spectrum and is mostly seen at relatively {\it low} photon energies (while electromagnetic corrections produce a sharp spectral feature at {\it high} $E_\gamma\sim m_\chi$). However, electroweak IB certainly induces a very sizable change of both  total photon flux and  spectral form. In particular, it is worth stressing that this is almost completely due to the full calculation presented here: using instead the often adopted model-independent expressions for FSR \cite{Ciafaloni:2010ti} -- as implemented, e.g., in Ref.~\cite{Cirelli:2010xx} -- results in a total photon flux for this model that is essentially indistinguishable from the tree-level result (dashed line).

\paragraph{Conclusions.---}

We have performed a full calculation of leading electroweak corrections to the 
annihilation cross section of supersymmetric neutralino DM, demonstrating that 
such corrections may significantly enhance the annihilation rate of WIMP 
DM in realistic particle physics frameworks. While these processes do not
produce pronounced spectral features in gamma rays like the corresponding 
electromagnetic corrections, they may -- depending on the detector threshold -- 
enhance the 
observationally also highly  relevant 
integrated photon yield by 
up to two orders of magnitude compared to the tree-level expectation.

Let us stress that previous results  available 
in the literature, both concerning 'model-independent' parameterizations of the photon 
yield \cite{Ciafaloni:2010ti,Cirelli:2010xx} and  analytic results under various limiting assumptions \cite{Bell:2010ei,Bell:2011eu,Bell:2011if,Garny:2011cj,Garny:2011ii,Ciafaloni:2011gv,Ciafaloni:2012gs},  are 
in many situations not reliable enough to reproduce be it the shape or 
the normalization of the gamma-ray spectrum. This clearly illustrates the need for 
detailed calculations as presented here, which we are convinced  will
prove important for future indirect DM searches. We refer to a companion 
paper \cite{fullanalysis} for a more detailed discussion, including 
implications for indirect DM searches with other cosmic ray species. The routines to
compute the processes discussed here will 
be fully available with 
the next public \ds~release.

\begin{table}[t!]
\begin{ruledtabular}
   \begin{tabular}{ccccc|ccccc}
    $m_0$ &  $m_{1/2}$    &   tan\,$\beta$  & $A_0$ & sgn  & $m_\chi$  & $Z_g$ &  $m_{\tilde\tau}$ & $\Omega h^2$\\
    {\scriptsize \mbox{[GeV]}} & {\scriptsize \mbox{[GeV]}} && {\scriptsize \mbox{[GeV]}}&$(\mu)$&  {\scriptsize \mbox{[GeV]}} & & {\scriptsize \mbox{[GeV]}}&\\[0.6ex]
    \hline
    &&&&&\\[-1.2ex]
   168  & 871 & 4.61 & -292 & -1 & 362.5 & 0.999 & 364.3 & 0.113\\
   \end{tabular}
\end{ruledtabular}
  \caption{cMSSM model parameters for  the spectrum shown in Fig.~\ref{fig:spectrum}, along with neutralino mass $m_\chi$, gaugino fraction $Z_g$, mass of the lightest stau $m_{\tilde\tau}$ and relic density $\Omega h^2$.}
 \label{tab_susy}
\end{table}

\medskip
\paragraph{Acknowledgments.} We warmly thank Paolo Gondolo for initiating us
so thoroughly and patiently into the secrets of the helicity amplitude calculations
performed in Refs.~\cite{Edsjo:1997bg,Edsjo:2003us}, and Bj\"orn Herrmann for 
providing detailed numerical cross-checks of our spectra and cross sections during 
an early stage of this work. We also gratefully acknowledge very useful conversations
with Mathias Garny and Michael Gustafsson.
T.B.~and F.C.~acknowledge support from the German Research Foundation 
(DFG) through the Emmy Noether grant BR 3954/1-1.

\end{document}